\documentstyle{article}

\begin{document}
\pagestyle{empty}
\centerline{\Large \bf Recursively Undecidable Properties of $N\hspace{-0.1cm}P$}
\vspace{\baselineskip}
\vspace{\baselineskip}
\centerline{Vladimir Naidenko}
\centerline{\small (Institute of Mathematics, Belarus}
\centerline{\small ul. Surganova 11, Minsk 220012, BELARUS}
\centerline{\small naidenko@im.bas-net.by)}

\vspace{\baselineskip}
\vspace{\baselineskip}
\vspace{\baselineskip}
\begin{sloppypar}
{\small
\noindent{\bf Abstract:} 
It is shown that there cannot be any algorithm that for a given 
nondeterministic polynomial-time Turing machine determinates 
whether or not the language recognized by this machine belongs to $P$.


\noindent{\bf Category:} F
} 

\vspace{\baselineskip}
\vspace{\baselineskip}
\noindent{\large\bf 1\quad Introduction}

\vspace{\baselineskip}
\noindent Arora [Arora 1994] defined four main complexity classes for $NP$-optimization
problems, and stated the following question. Is there a method 
(at least an intuitive level) for recognizing, for a given problem,
which of these classes it fall in ? Kolaitis and Thakur [Kolaitis and Thakur 1995]
show that,
assuming that $NP\neq P$, it is an undecidable problem to tell
if a given first-order formula defines an approximable 
$NP$-optimization problem. We will prove a more general result that
there cannot be any algorithm which determinates for a given 
nondeterministic polynomial-time Turing machine 
whether or not the language recognized by this machine belongs to $P$.

\vspace{\baselineskip}
\vspace{\baselineskip}
\noindent{\large \bf 2\quad Main Results}

\vspace{\baselineskip}
\noindent Let $\Sigma $ be a fixed alphabet which contains at least two
symbols, $\#$\ a symbol not being in $\Sigma $, and $L_{NP}$ a fixed $NP$-complete
language over the alphabet $\Sigma $. By $G$ and $L(G)$ we mean the
context-free grammar and language respectively. Let $\propto $ be a
metasymbol of polynomial-time Turing reducability.
\par
With every context-free grammar $G=(V,\Sigma ,P,\sigma )$ with the
terminal alphabet $\Sigma$ ([see Ginsburg 1966]), we associate the language
$(\Sigma ^{*}\backslash L(G))\#L_{NP}$, i.e. the language $(\Sigma ^{*}\backslash L(G))\#L_{NP}$ is a
concatenation of the languages $\Sigma ^{*}\backslash L(G), \{\#\}$ and $L_{NP}$ :
$$
(\Sigma ^{*}\backslash L(G))\#L_{NP}\hbox{ = }\{ x\#y \mid x \in \Sigma ^{*}\backslash L(G) , y \in L_{NP} , \# \not\in \Sigma \} .
$$
For languages of the form $(\Sigma ^{*}\backslash L(G))\#L_{NP}$ the following
theorems hold.
\par
{\it Theorem 1}. Any language of the form $(\Sigma ^{*}\backslash L(G))\#L_{NP}$ belongs
to $NP$.
\par
{\it Proof} folows from the fact that any context-free language
$L(G)$ belongs to $P$. $\Box$
\par
{\it Theorem 2}. Provided that $NP\neq P$, the language $(\Sigma ^{*}\backslash L(G))\#L_{NP}$
belongs to $P$ iff $L(G) = \Sigma ^{*}$.
\par
{\it Proof}. Suppose $L(G) \neq \Sigma ^{*}$. Then $L_{NP} \propto (\Sigma ^{*}\backslash L(G))\#L_{NP}$ .
Therefore, provided that $NP \neq P$,\ it follows
$(\Sigma ^{*}\backslash L(G))\#L_{NP} \not\in P$.
Suppose $L(G) = \Sigma ^{*}$. Then $(\Sigma ^{*}\backslash L(G))\#L_{NP} = \emptyset $ . Consequently,
$(\Sigma ^{*}\backslash L(G))\#L_{NP} \in P$. $\Box$
\par
{\it Theorem 3}. There cannot be any algorithm that for a given
nondeterministic polynomial-time Turing machine determinates
whether or not the language recognized by this machine belongs to $P$.
\par
{\it Proof}. Suppose that such an algorithm is found. Then one
can determinate for a Turing machine accepting the language of
the form $(\Sigma ^{*}\backslash L(G))\#L_{NP}$ whether or not
this language belongs to $P$. 
Note that such a Turing machine can be effectively
constructed from a given context-free grammar $G$. Therefore, if
the algorithm tests membership of $(\Sigma ^{*}\backslash L(G))\#L_{NP}$ in $P$ then using
its output one can verify the truth of assertion $L(G) = \Sigma ^{*}$ for a
given context-free grammar $G$ (by theorem 2) . However, this
is impossible because of the recursive unsolvability of
assertion $L(G) = \Sigma ^{*}$ for an arbitrary context-free grammar $G$.
$\Box$
\par
{\it Theorem 4}. The following question is undecidable:
\par
"Is the language $NP$-complete, accepted by a given
nondeterministic polynomial-time Turing machine ?"
\par
{\it Proof} follows from theorem 3. $\Box$
\par
Let us build other languages from $NP$, possessing the same
undecidable propeties as the language $(\Sigma ^{*}\backslash L(G))\#L_{NP}$ .
\par
Denote the set of all subsets of $\Sigma ^{*}$ by $2^{{\Sigma}^{*}}$. Introduce a
function $f$ mapping $2^{{\Sigma}^{*}}$ into $2^{{\Sigma}^{*}}$ and defined by the following
equation
\par
\noindent $f(L) = \{ x \in \Sigma ^{*} \mid $ there exists a chain (word) $w$ such that
$\mid w\mid \le \mid x\mid $ and $w \in \Sigma ^{*}\backslash L \}$,
\par
\noindent where $L$ is an argument taking the values from $2^{{\Sigma}^{*}}$.
\par
If one substitutes a context-free language $L(G) \subseteq \Sigma ^{*}$ for the
argument $L$ then the language $f(L(G)) \subseteq \Sigma ^{*}$ is obtained.
\par
It is clear that any language of the form $f(L(G)) \cap L_{NP}$
belongs to $NP$.
\par
{\it Theorem 5}. Provided that $P \neq NP$, the language $f(L(G)) \cap L_{NP}$
belongs to $P$ iff $L(G) = \Sigma ^{*}$.
\par
{\it Proof} is obvious. $\Box$
\par
Let us consider the language $f(L(G)) \cup L_{NP}$ . It possesses
the analogical undecidable propeties. Clearly, $f(L(G)) \cup L_{NP} \in NP$.
If $L(G) = \Sigma ^{*}$ then $f(L(G)) \cup L_{NP} = L_{NP}$. Consequently, the language
$f(L(G)) \cup L_{NP}$ doesn't belong to $P$, provided that $L(G) = \Sigma ^{*}$. If
$L(G) \neq \Sigma ^{*}$ then the language is co-finite. Therefore,
$f(L(G)) \cup L_{NP} \in P$ when $L(G) \neq \Sigma ^{*}$.

\newpage
\noindent{\large\bf References}
\vspace{\baselineskip}

{\small
\noindent [Arora 1994] Arora, S.: "Probabilistic checking of proofs and 
hardness of approximation \linebreak\indent
problems"; Technical report CS-TR-476-94, 
Princeton, USA (1994).

\noindent [Kolaitis and Thakur 1995] Kolaitis, Ph., Thakur, M.:
"Approximation propeties of \linebreak\indent
NP minimization classes"; J.Comput.System Sci. 50, 3 (1995), 391-411.
\par
\noindent [Ginsburg 1966] Ginsburg, S.: "The mathematical 
theory of context-free
languages"; \linebreak\indent
McGraw-hill Book Company, Inc.\,/\,New York (1966).
\par
}
\end{sloppypar}
\end{document}